\documentclass[a4paper]{article}

\usepackage{INTERSPEECH2022}
\usepackage{xcolor}
\usepackage{booktabs}
\usepackage{subcaption}
\usepackage{multirow}
\usepackage{etoolbox}
\usepackage{url}
\usepackage{hyperref}

\apptocmd{\thebibliography}{\setlength{\itemsep}{-0mm}\setlength{\baselineskip}{-0mm}}{}{}

    \title{Improving Speech Enhancement through Fine-Grained Speech Characteristics}
\name{Muqiao Yang$^{1\dagger}$, Joseph Konan$^{1\dagger}$, David Bick$^{1\dagger}$, Anurag Kumar$^2$, Shinji Watanabe$^1$, Bhiksha Raj$^{1}$ \thanks{$^\dagger$ Equal contribution. Reverse alphabetical order.} }
\address{
  $^1$ Carnegie Mellon University\\
  $^2$ Meta Reality Labs Research
}
\email{ \{muqiaoy, jkonan, dbick, bhiksha\}@cs.cmu.edu, swatanab@andrew.cmu.edu, anuragkr@fb.com}

\begin{document}

\graphicspath{{images/}}
\maketitle
\begin{abstract}
  While deep learning based speech enhancement systems have made rapid progress in improving the quality of speech signals, they can still produce outputs that contain artifacts and can sound unnatural. We propose a novel approach to speech enhancement aimed at improving perceptual quality and naturalness of enhanced signals by optimizing for key characteristics of speech. We first identify key acoustic parameters that have been found to correlate well with voice quality (e.g. jitter, shimmer, and spectral flux) and then propose objective functions which are aimed at reducing the difference between clean speech and enhanced speech with respect to these features. The full set of acoustic features is the extended Geneva Acoustic Parameter Set (eGeMAPS), which includes 25 different attributes associated with perception of speech. Given the non-differentiable nature of these feature computation, we first build differentiable estimators of the eGeMAPS and then use them to fine-tune existing speech enhancement systems. Our approach is generic and can be applied to any existing deep learning based enhancement systems to further improve the enhanced speech signals. Experimental results conducted on the Deep Noise Suppression (DNS) Challenge dataset shows that our approach can improve the state-of-the-art deep learning based enhancement systems. 
  
\end{abstract}

\noindent\textbf{Index Terms}: speech enhancement, eGeMAPS, acoustic parameters, perceptual quality, explainable enhancement evaluation.

\section{Introduction}

Speech enhancement (SE) is aimed at improving the quality and intelligibility of speech signals that have been degraded by noise signals, reverberations and other factors. The need for speech enhancement arises in a variety of applications; communications through VoIP, mobile phones, in hearing aids, in downstream applications such as automatic speech recognition, and many others. A large body of SE literature focuses on single-channel speech enhancement and our focus in this paper is also on single-channel SE. 


Classical approaches for SE have been signal processing driven. These include methods such as spectral subtraction and Wiener filtering \cite{spectral_subraction, weiner_filtering_denoising}. However, much of the recent progress in SE comes from deep learning methods. A variety of deep neural network (DNN) based speech enhancement methods have been proposed and improvements have been made through every subsequent architecture \cite{Lu2013SpeechEB, regression_dnn_denoising, rnn_denoising, Weninger2015SpeechEW, cnn_denoising, segan}. A key component of these approaches is the loss function used for training the neural networks. Once again a wide variety of methods have been proposed, but generally these loss functions try to minimize $L_1$/$L_2$ loss between the network output and the target clean speech, sometimes in time-domain \cite{defossez2020real}, sometimes in time-frequency domain \cite{time_frequency_domain}. 
However, these point-wise differences fail to capture key speech characteristics which can lead to enhanced speech with artifacts and unnatural speech \cite{dnsmos_pesq_flaw}. For example, these flaws can arise due to failure to capture pitch differences \cite{turian_henry} and ignoring low-energy regions that represent phonemes such as fricatives or plosives \cite{phoneme_classification_loss}. 

The aforementioned problems led to recent speech enhancement research that specifically optimizes for perceptual quality of the signals. One branch of these works uses differentiable estimators of existing perceptual metrics to optimize their model \cite{fu2019metricgan, metric_RL_SE, white_box_perceptual_loss}, including Perceptual Evaluation of Speech Quality (PESQ) \cite{PESQ} and Short-Time Objective Intelligibility (STOI) \cite{STOI}. However, these objective metrics correlate well with human perception only to a certain degree \cite{reddy2021dnsmos, dnsmos_pesq_flaw, limitations_of_pesq} and the benefits of optimizing these metrics are often observed to be limited \cite{black_box_peq_stoi, metric_RL_SE, other_metric_approach}.


Another class of works in this direction does not target any specific metric but attempts to capture perceptual quality through auxiliary losses. Common approaches involve using differences between filter banks or deep representations \cite{saddler2020speech, pfpl_paper, seven_ensemble_model}. However, these losses also provide only implicit supervision and do not specifically target speech characteristics, leading to limited improvements. 


In this paper, we argue that optimizing many specific acoustic qualities - the fine-grained features - of enhanced speech will result in improved perceptual quality. These features are important because they have shown to correlate with perceptual quality, such as in breathy or harsh voice quality \cite{correlates_breathy,  correlates_breathy_rough}, and raspy voice quality \cite{KASUYA1986171}. They have even shown correlation with perception of a speaker's personality, such as warmth \cite{Gallardo2017PerceivedIS}. We show that two state-of-the-art models produced enhanced speech with different values for these features than clean speech. This shows the potential of our method compared to metric or other loss optimization.

Unlike any prior works, we explicitly optimize for these important acoustic properties and force the network to pay attention to these low-level features. We use the extended Geneva Minimal Parameter Set (eGeMAPS) \cite{egemaps} as the full feature set, for its coverage of many categories of speech information. Every feature was selected for its association with cues of voice quality or emotion. Existing calculations for these features are non-differentiable, so we train an eGeMAPS prediction network that can plug into other learning pipelines. Its effectiveness is demonstrated by mean-squared error (MSE) of predicted and ground-truth eGeMAPS. After 50 epochs, we reach MSE of $1.9 \cdot 10^{-4}$ on train and $2.6 \cdot 10^{-4}$ on test.

We verify our hypothesis experimentally, by fine-tuning two existing speech enhancement systems with our auxiliary loss to minimize the difference between eGeMAPS features for enhanced and clean speech. Specifically, we chose two models that rate highly on standard metrics, FullSubNet\footnote{\url{https://github.com/haoxiangsnr/FullSubNet}} \cite{hao2021fullsubnet} and Demucs\footnote{\url{https://github.com/facebookresearch/denoiser}} \cite{defossez2020real}. Both were trained on data from the Deep Noise Suppression (DNS) Challenge of InterSpeech 2020 \cite{reddy2020interspeech}, and provided pre-trained models. We initialize from these checkpoints, and fine-tune on this dataset because of its large size and variety of noise-types. Our experimental results on these systems show improvements in widely used metrics, PESQ, STOI and MOS (DNSMOS \cite{reddy2021dnsmos} as proxy for MOS). 


\section{Related Work}
Our approach falls into the space of speech enhancement methods that are aimed for improving perceptual quality. There are two primary classes of works in this area: (a) objective metric optimization, and (b) feature loss optimization. 


\subsection{Objective Metric Optimization}

One recent approach targeting a metric used the generative adversarial network (GAN) framework. The discriminator is a differentiable PESQ estimator, and the generator is the enhancement model \cite{fu2021metricgan, fu2019metricgan}. Other approaches use the metrics as rewards within a reinforcement learning context \cite{metric_RL_SE}, or formulate a differentiable equation that approximates the metric \cite{white_box_perceptual_loss}. However, it is hard to obtain strong improvements through implicit supervision from a collapsed metric. Similar to these other methods, we use a neural network as a differentiable estimator of previously non-differentiable quantities. However, unlike these works, we explicitly optimize for natural speech characteristics captured through a set of low-level acoustic features.

  
\subsection{Feature Losses} 
This class of works cover a broader spectrum but they generally try to incorporate domain-knowledge in forms of feature-based loss functions. For example, \cite{saddler2020speech} attempts to bring domain-knowledge from computational models of human auditory perception by minimizing the difference between filter-banks meant to stimulate human cochlea \cite{saddler2020speech}.  \cite{phoneme_classification_loss} encodes linguistic domain-knowledge by matching senone logits between clean and enhanced speech. \cite{seven_ensemble_model} injects emotion and speaker related information. Some have also argued for using deep speech representations for improving enhancement models \cite{pfpl_paper}. Among these methods, our method is along the lines of the first two approaches, as we also target specific features of speech. However, our approach is very different from these as we rely on differentiable estimators of well-defined acoustic features in eGeMAPS set to capture speech characteristics.



\section{Proposed Framework}

\subsection{Acoustic Parameters}
The eGeMAPS is a collection of acoustic parameters that provide details about an audio signal \cite{egemaps}. For example, consider \emph{jitter} features. A perfect sinusoid has a consistent period, but the period length for a speech signal varies by milliseconds (ms) from period to period. The difference between these durations for the fundamental frequency (F0) is defined as \emph{jitter}. Similarly, the amplitude reaches slightly different heights for each period. Shimmer is the difference between peak amplitudes for consecutive periods of F0. 

\begin{figure}[t]
    \centering
    \includegraphics[width=\columnwidth]{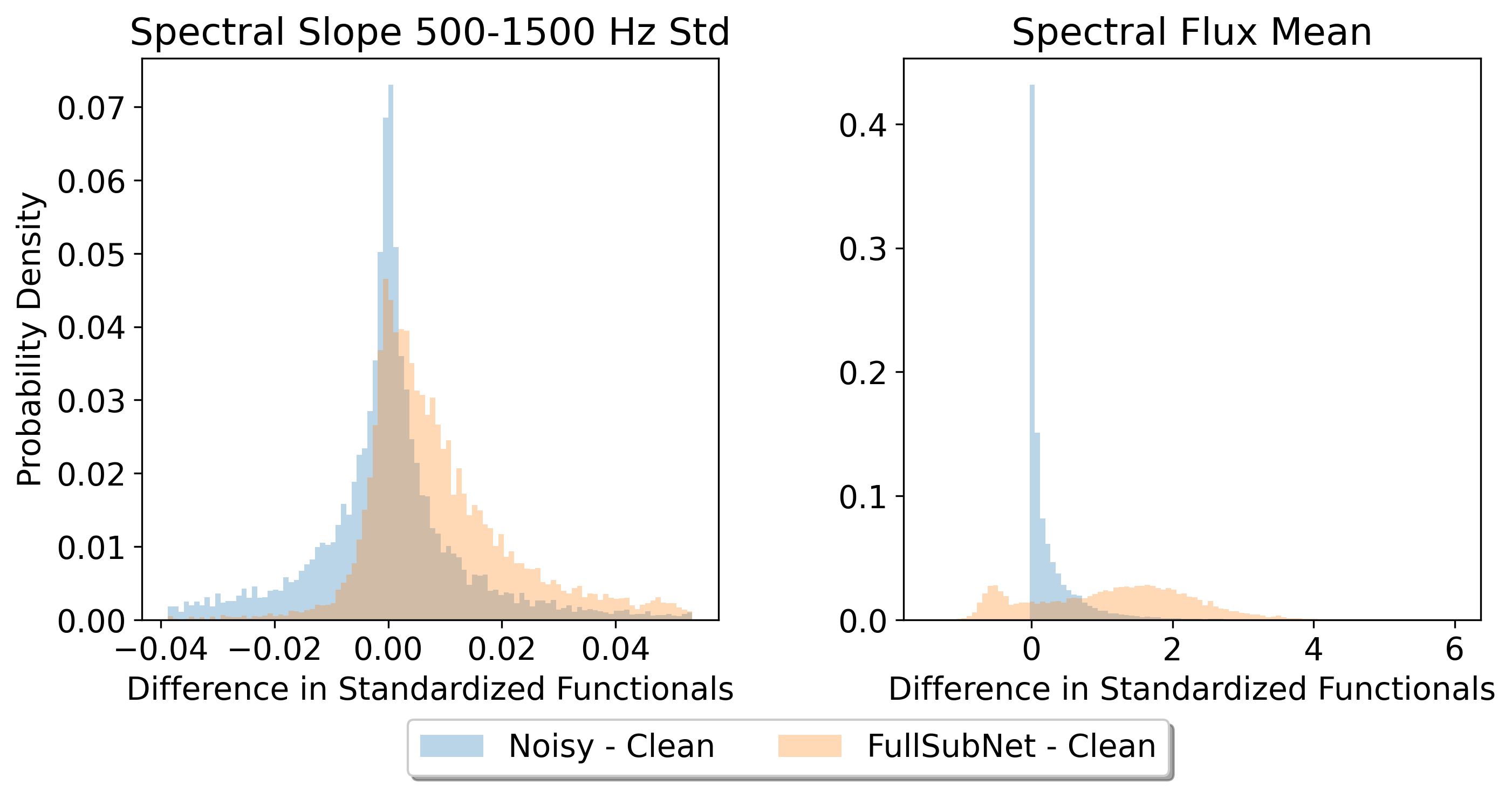}
    \caption{Example distributions of differences between standardized functionals. In both plots we see differences between enhanced and clean speech, with the right plot showing greater differences than noisy speech.}
    \label{fig:enhancement-comp}
    \vspace{-4mm}
\end{figure}

Researchers have discovered that jitter and shimmer correlate with breathy or harsh voice quality \cite{correlates_breathy, KASUYA1986171}. Another acoustic parameter, harmonics-to-noise ratio, correlates well with raspy vocal quality \cite{correlates_breathy_rough}. Spectral flux and spectral slope have been correlated with the perception of an individual's personality, such as their warmth  \cite{Gallardo2017PerceivedIS}. Each of the 25 eGeMAPS were chosen by speech experts based on literature associating them with the perception of voice \cite{egemaps}.

Given this association of acoustic parameters and perception of voice, we hypothesize that the eGeMAPS could help us quantify the perceptual quality of enhanced speech. We also hypothesize that inducing similarities in these features could improve the quality of the enhancement.

The eGeMAPS are produced in a two-step process. Initially, the 25 features are calculated for windows of 30 ms, with a stride of 10 ms. These are called low-level descriptors (LLDs). Next, statistics are calculated over LLDs, including mean, standard deviation, and percentiles. These are called functionals, and eGeMAPS defines 88 functionals per input. Let $D = 88$ denote the size of each eGeMAPS functionals vector for an utterance.

\begin{figure}
    \centering
    \includegraphics[width=\columnwidth]{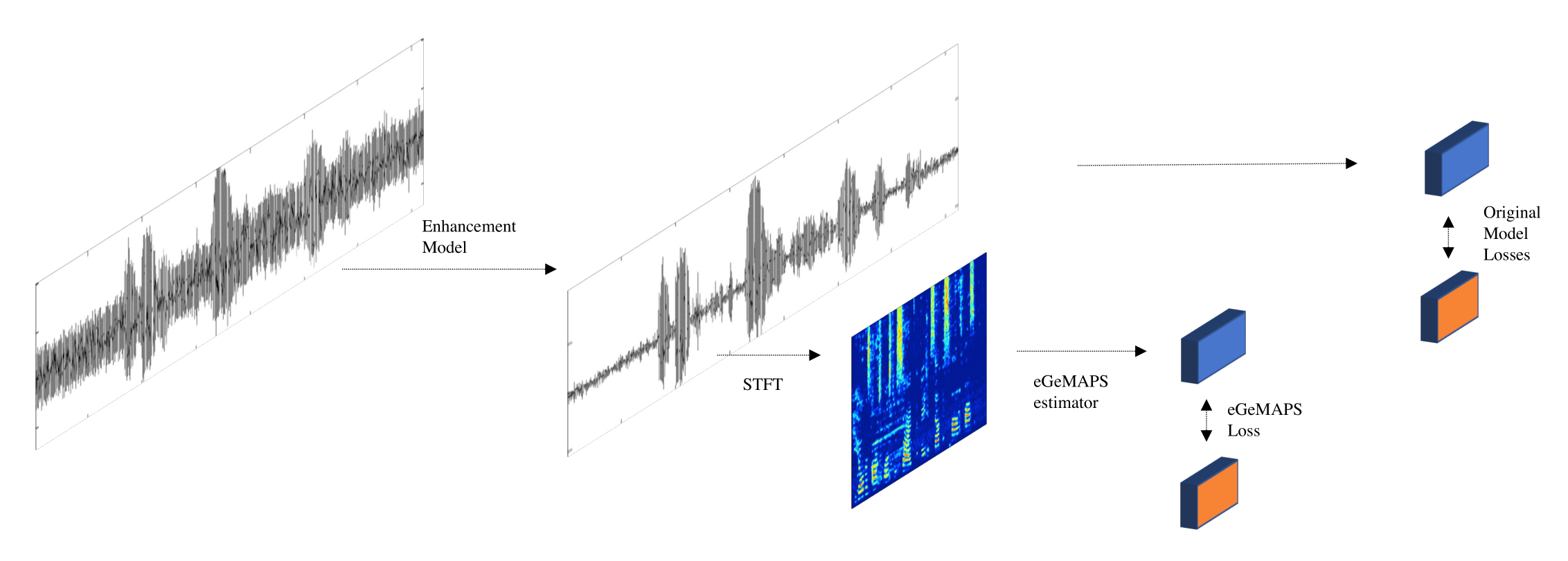}
    \caption{Our fine-tuning pipeline for arbitrary model. Our proposed loss is on the left. Note that the original model losses can be one or many, and in time domain, frequency domain, or other.}
    \label{fig:diagram}
    \vspace{-3mm}
\end{figure}

\begin{figure*}
    \centering
    \includegraphics[width=\textwidth]{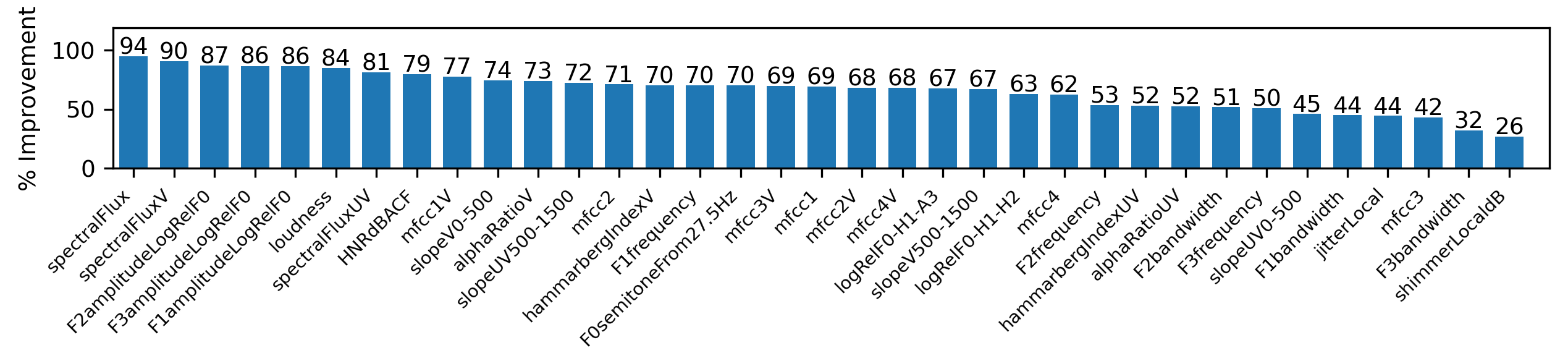}
    \caption{Percent improvement of the eGeMAPS functionals before and after our fine-tuning. Percent improvement is measured in reduction of difference from clean speech. Labels of openSMILE\cite{opensmile} features are included to allow searching of feature names.}
    \label{fig:each-ege-improvement}
    \vspace{-3mm}
\end{figure*}

\subsection{Visualization of eGeMAPS Differences}
\label{sec:viz}
First, we show that enhanced speech differs from clean across many features. In Figure \ref{fig:enhancement-comp} we plot the differences in eGeMAPS functionals of clean speech with both enhanced and noisy speech \cite{egemaps}. Since the functionals have different scales, to accurately reflect the scale of differences in the plots, we standardize each functional according to the mean and standard deviation (std) of the \textit{clean} functional values.

For each utterance in the data, indexed by $n \in 1, \dots, N$, we have a vector of differences $\mathbf{d}^{(n)} \in \mathbb{R}^{D}$. Each $\mathbf{d}^{(n)}_{i}, i \in \{1, \dots, D\}$ is the difference between clean and enhanced/noisy for eGeMAPS functional $i$. Figure \ref{fig:enhancement-comp} shows $\mathbf{d}^{(n)} \: \forall n \in \{1, \dots, N\}$ for two functionals: the standard deviation of spectral slope and the mean of spectral flux. The x-axis is the difference in standardized values. The y-axis is the probability of an observation falling in a bin of x-values. 

The ideal distribution for eGeMAPS differences of enhanced and clean speech would look close to a Dirac delta function with all mass centered at 0. It would show enhanced speech with matching acoustic features to clean, which we hypothesize provides high perceptual quality. When evaluating the model, any mean different from 0 shows worse performance, and higher variance is worse because it shows more differences of higher values. 

In Figure \ref{fig:enhancement-comp}, we see that enhanced speech differs from clean speech in a substantial way. The right-hand side shows that enhanced speech differs even more than noisy speech. These large differences show potential for our optimization to induce closer feature values for enhanced and clean speech and improve perceptual quality and naturalness of speech.

\subsection{Modeling}
\label{sec:modeling}
We now describe our modeling approach to reduce the difference between eGeMAPS for enhanced and clean speech. We first describe each model's inputs and outputs, and how these plug into our method.   

Demucs takes in the noisy time-domain signal $\mathbf{x} \in \mathbb{R}^L, \mathbf{x} = \mathbf{y} + \mathbf{n}$, where $\mathbf{y} \in \mathbb{R}^L$ is the clean waveform and $\mathbf{n} \in \mathbb{R}^L$ is the noise waveform. It directly outputs the enhanced waveform $\mathbf{\hat{y}} \in \mathbb{R}^L, \mathbf{\hat{y}} = g_{\mathbf{\theta}}(\mathbf{x})$. Let $g$ be the speech enhancement model parameterized by weights $\mathbf{\theta}$, in this case Demucs. The weights are trained to minimize the $L_1$ loss between clean and enhanced waveforms. It also supplements with a loss between spectrograms, calculated by the short-time Fourier transform (STFT). For further details on the architecture we defer to the original paper \cite{defossez2020real}. 

FullSubNet is composed of two models that operate on different parts of the noisy magnitude spectrogram $\mathbf{X} \in \mathbb{R}^{T \times F}$, which is the real-valued output of the short-time Fourier transform over $\mathbf{x}$. The model trains to minimize the $L_1$ or $L_2$ loss between complex Ratio Mask $\mathbf{M} \in \mathbb{R}^{T \times F}$ and the target complex Ideal Ratio Mask \cite{complex_ratio_mask}. The mask is multiplied element-wise with the noisy spectrogram to obtain enhanced spectrogram $\mathbf{\hat{Y}} = \mathbf{M} \cdot \mathbf{X}$. Finally, inverse-STFT (iSTFT) gives us the desired waveform $\mathbf{\hat{y}}$. 


For each model, we supplement the existing loss with our proposed eGeMAPS loss. Our estimator produces the eGeMAPS estimates from the enhanced spectrogram: 

\begin{equation}
\mathbf{\hat{e}} = h_{\phi}(\mathbf{\hat{Y}}), \; \mathbf{\hat{e}} \in \mathbb{R}^D
\end{equation}
where $h_{\phi}$ is our estimator parameterized by weights $\mathbf{\phi}$. We apply STFT on $\mathbf{\hat{y}}$ to get $\mathbf{\hat{Y}}$, which allows us to apply our method to any enhancement model.

We also define the original non-differentiable function to calculate eGeMAPS given by \cite{egemaps}, which is implemented in the openSMILE\cite{opensmile} toolkit:
\begin{equation}
\mathbf{e} = o(\mathbf{y}), \mathbf{e} \in \mathbb{R}^D
\end{equation}

The toolkit takes in waveform $\mathbf{y}$ rather than spectrogram $\mathbf{Y}$. Given the difficulty of modeling waveforms, our estimator operates on spectrograms. Our training results for estimating eGeMAPS from spectrograms validate this decision. 

For each batch of size $B$, our loss is the mean squared error between estimated eGeMAPS from enhanced speech and ground-truth eGeMAPS from clean speech: 
\begin{equation}
\mathcal{L}^{\text{eGeMAPS}}_{\theta, \phi} = \frac{1}{B} \sum_{i=1}^{B} \lVert \mathbf{\hat{e}}^{\mathrm{enh}}_i - \mathbf{e_i}^{\mathrm{clean}} \rVert^2_2
\end{equation} 
where superscripts $\mathrm{enh}$ and $\mathrm{clean}$ denote enhanced and clean speech, respectively. We combine each model's original loss with our loss: 
\begin{equation}
\mathcal{L}^{\text{final}}_{\theta, \phi} = \mathcal{L}^{\text{original}}_{\theta} + \lambda \mathcal{L}^{\text{eGeMAPS}}_{\theta, \phi}
\end{equation}

Note that $\mathcal{L}^{\text{eGeMAPS}}$ depends on both the enhancement model parameters $\theta$ \textit{and} estimator parameters $\phi$, while the original SE model objectives only depend on $\theta$. Based on $\mathcal{L}^{\text{final}}_{\theta, \phi}$ we fine-tune the enhancement weights $\theta$ to optimize the original and eGeMAPS objectives. We also experiment with fixing and fine-tuning $\phi$, which we discuss further in Section \ref{sec:ablation}. Figure \ref{fig:diagram} visualizes this fine-tuning scheme for arbitrary enhancement models.

\section{Experiments}
\subsection{Data}
We used Deep Noise Suppression (DNS) Challenge 2020 \cite{reddy2020interspeech} data synthesized from their provided script, which creates 12,000 utterances of 30 seconds (s). We synthesized another 1,200 utterances for a validation set. The clean samples are from Librivox\footnote{\url{https://librivox.org/}}, and added noise clips are from Audioset\footnote{\url{https://research.google.com/audioset/}} and Freesound\footnote{\url{https://freesound.org/}}. The Signal to Noise Ratio (SNR) is sampled uniformly between 0 and 40 decibels (dB). Demucs and FullSubnet both train on sub-samples of the 30s audios. We followed Demucs' configuration: for each 30 second audio, we took four seconds of audio with one second stride to create 27 samples.  
We used the DNS synthetic test set with no reverberation. This set consists of 150 utterances from Graz University's clean speech dataset \cite{graz_university}, combined with noise categories chosen to represent real-world noise interruptions to audio calls. The SNR was sampled uniformly between 0 and 25 dB.

\subsection{Estimator Specifications}
We obtained our eGeMAPS estimator by first pre-training a Predictive Aux-VAE \cite{springenberg19_interspeech} on the clean spectrograms of our dataset. Pre-training provides quality representations of spectrograms and allows for easier optimization of downstream prediction of eGeMAPS. We removed the original VAE decoder after spectrogram pre-training. For our final eGeMAPS estimator, we used the VAE encoder followed by self-attention pooling \cite{self_attention_pooling} and two linear layers. The encoder is composed of four layers of convolutions, with the first three followed by tanh non-linearity and batch-normalization \cite{ioffe2015batch}. We confirm the estimator is performant by showing MSE values of $1.9 \cdot 10^{-4}$, $1.6 \cdot 10^{-4}$, and $2.6 \cdot 10^{-4}$ on train, validation, and test. Code for the estimator and fine-tuning of enhancement will be available at \textcolor{blue}{\url{https://github.com/muqiaoy/eGeMAPS_estimator}}.

\begin{figure}[t]
    \centering
    \includegraphics[width=\columnwidth]{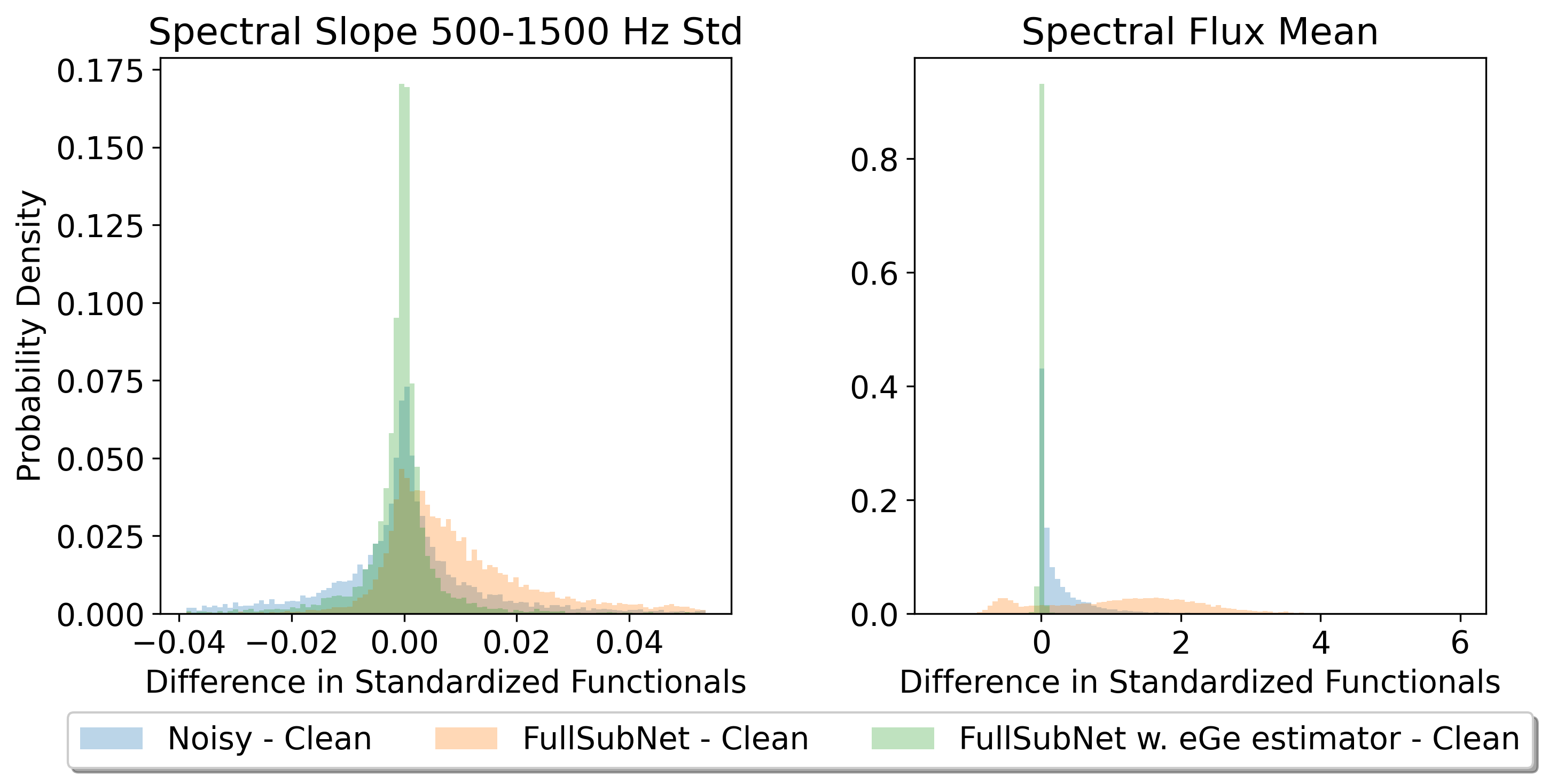}
    \caption{Test data distribution of eGeMAPS differences after fine-tuning, showing that we have pushed the differences towards 0. In the right-hand figure, the fine-tuning corrects the feature that was worse than noisy speech.}
    \label{fig:fine-tuning-success}
\end{figure}

\begin{table}[t]
    \centering
    \begin{tabular}{c c c c}
    \toprule
    Model & PESQ & STOI & DNSMOS \\
    \midrule
    Clean & - & - & 4.02 \\
    Noisy & 1.58 & 91.52 & 3.16 \\
    \midrule
        FullSubNet \cite{hao2021fullsubnet} & 2.89 & 96.41 & 3.91 \\
        \multirow{1}{*}{FullSubNet +} & \multirow{2}{*}{\textbf{2.91}} & \multirow{2}{*}{\textbf{96.43}} & \multirow{2}{*}{\textbf{3.93}} \\
        \multirow{1}{*}{eGeMAPS estimator} \\
    \midrule
        Demucs \cite{defossez2020real} & 2.65 & 96.54 & \textbf{3.80} \\
        \multirow{1}{*}{Demucs +} & \multirow{2}{*}{\textbf{2.81}} & \multirow{2}{*}{\textbf{96.83}} & \multirow{2}{*}{3.78} \\
        \multirow{1}{*}{eGeMAPS estimator} \\
    \bottomrule
    \end{tabular}
    \caption{Experimental results showing the benefit of eGeMAPS loss on FullSubNet and Demucs, measured by standard perceptual metrics. PESQ and STOI for clean are blank because the metrics are calculated with clean as reference.}
    \label{tab:fine-tune-results-test}
    \vspace{-5mm}
    
\end{table}

\subsection{Results}
In Figure \ref{fig:each-ege-improvement} we see the success of our fine-tuning. Recall that eGeMAPS functionals are statistics of LLDs, such as mean and std. We show percent improvement in the means, where percent improvement is the reduction in difference from clean speech. Therefore 100\% improvement means that the difference has been removed, and the functional matches clean. We see improvement in all features.

Figure \ref{fig:fine-tuning-success} provides another way to visualize this improvement, through plotting differences between eGeMAPS as in Section \ref{sec:viz}. The green distribution on the right-hand side of the plot now shows close to ideal distribution of differences, completely peaked around 0. The left-hand side does not fully minimize the differences but shows good improvements compared to the original enhancement.

In Table \ref{tab:fine-tune-results-test} we show the effect of improvements in eGeMAPS differences on existing perceptual metrics. Note that PESQ and STOI are calculated with clean audios as reference, while DNSMOS \cite{reddy2021dnsmos} is directly evaluated on the enhanced audios. 
We report DNSMOS on clean to show our upper bound of performance. We see modest improvements in all metrics for FullSubNet. We also show large improvements in PESQ and STOI for Demucs. These indicate that our eGeMAPS estimator is improving the overall speech quality by the fine-grained speech characteristics.


\subsection{Ablation Study of eGeMAPS Estimator}
\label{sec:ablation}
We analyzed multiple configurations for pairing the estimator with enhancement models, shown in Table \ref{tab:ablation}. In all configurations, the original VAE encoder weights were fixed. We experimented with pre-training the final linear layers before fine-tuning the speech-enhancement model. We also tried training them while fine-tuning the enhancement model, but this did not converge. We found that fine-tuning the estimator along with the enhancement model performed worse in initial experiments, and therefore decided to perform our hyperparameter search with fixed estimator. We hypothesized that fine-tuning the estimator could add robustness to enhanced speech as input, but we conjecture that it creates a harder optimization problem. 

We experimented with the weight multiplier, $\lambda$, of our eGeMAPS loss to match the scale of loss values, but found that $\lambda = 1$ is optimal for fine-tuning $\mathbf{\theta}$ when using fixed $\mathbf{\phi}$.

\begin{table}
    \centering
    \begin{tabular}{c c c c }
    \toprule
        Setting & PESQ & STOI \\
    \midrule
        $\lambda=0$ & 2.65 & 96.54 \\
    \midrule
        $\lambda=0.1$ & 2.78 & 96.77 \\
        $\lambda=1$ & \textbf{2.81} & \textbf{96.83} \\
        $\lambda=10$ & 2.80 & 96.75 \\
    \midrule
        $\lambda=1$, fine-tuning $\phi$ & 2.79 & 96.75 \\
        
    \bottomrule 
    \end{tabular}
    \caption{An ablation study of the effect of eGeMAPS estimator on Demucs, where $\phi$ refers to the estimator weights as introduced in Section \ref{sec:modeling}. If not specified, $\phi$ is fixed during the enhancement model fine-tuning stage.}
    \label{tab:ablation}
    \vspace{-5mm}
\end{table}

\section{Conclusion}
We demonstrate that fine-grained speech characteristics can significantly improve speech enhancement. We create a differentiable eGeMAPS estimator, allowing us to fine-tune existing models by minimizing acoustic parameter differences. The resulting enhanced audio yields superior results over traditional methods that marginally improve acoustic features and sometimes make them worse. Most perceptual evaluation metrics and acoustic functional statistics can be improved using our approach.


\bibliographystyle{IEEEtran}

\clearpage
\bibliography{mybib}

\begin{thebibliography}{10}
\providecommand{\url}[1]{#1}
\csname url@samestyle\endcsname
\providecommand{\newblock}{\relax}
\providecommand{\bibinfo}[2]{#2}
\providecommand{\BIBentrySTDinterwordspacing}{\spaceskip=0pt\relax}
\providecommand{\BIBentryALTinterwordstretchfactor}{4}
\providecommand{\BIBentryALTinterwordspacing}{\spaceskip=\fontdimen2\font plus
\BIBentryALTinterwordstretchfactor\fontdimen3\font minus
  \fontdimen4\font\relax}
\providecommand{\BIBforeignlanguage}[2]{{%
\expandafter\ifx\csname l@#1\endcsname\relax
\typeout{** WARNING: IEEEtran.bst: No hyphenation pattern has been}%
\typeout{** loaded for the language `#1'. Using the pattern for}%
\typeout{** the default language instead.}%
\else
\language=\csname l@#1\endcsname
\fi
#2}}
\providecommand{\BIBdecl}{\relax}
\BIBdecl

\bibitem{spectral_subraction}
S.~Boll, ``Suppression of acoustic noise in speech using spectral
  subtraction,'' \emph{IEEE Transactions on Acoustics, Speech, and Signal
  Processing}, vol.~27, no.~2, pp. 113--120, 1979.

\bibitem{weiner_filtering_denoising}
P.~Scalart and J.~Filho, ``Speech enhancement based on a priori signal to noise
  estimation,'' in \emph{1996 IEEE International Conference on Acoustics,
  Speech, and Signal Processing Conference Proceedings}, vol.~2, 1996, pp.
  629--632 vol. 2.

\bibitem{Lu2013SpeechEB}
X.~Lu, Y.~Tsao, S.~Matsuda, and C.~Hori, ``Speech enhancement based on deep
  denoising autoencoder,'' in \emph{INTERSPEECH}, 2013.

\bibitem{regression_dnn_denoising}
Y.~Xu, J.~Du, L.-R. Dai, and C.-H. Lee, ``A regression approach to speech
  enhancement based on deep neural networks,'' \emph{IEEE/ACM Transactions on
  Audio, Speech, and Language Processing}, vol.~23, no.~1, pp. 7--19, 2015.

\bibitem{rnn_denoising}
F.~Weninger, F.~Eyben, and B.~Schuller, ``Single-channel speech separation with
  memory-enhanced recurrent neural networks,'' in \emph{2014 IEEE International
  Conference on Acoustics, Speech and Signal Processing (ICASSP)}, 2014, pp.
  3709--3713.

\bibitem{Weninger2015SpeechEW}
F.~Weninger, H.~Erdogan, S.~Watanabe, E.~Vincent, J.~L. Roux, J.~R. Hershey,
  and B.~Schuller, ``Speech enhancement with {LSTM} recurrent neural networks
  and its application to noise-robust {ASR},'' in \emph{LVA/ICA}, 2015.

\bibitem{cnn_denoising}
\BIBentryALTinterwordspacing
H.~Zhao, S.~Zarar, I.~Tashev, and C.~Lee, ``Convolutional-recurrent neural
  networks for speech enhancement,'' \emph{CoRR}, vol. abs/1805.00579, 2018.
  [Online]. Available: \url{http://arxiv.org/abs/1805.00579}
\BIBentrySTDinterwordspacing

\bibitem{segan}
\BIBentryALTinterwordspacing
S.~Pascual, A.~Bonafonte, and J.~Serr{\`{a}}, ``{SEGAN:} speech enhancement
  generative adversarial network,'' \emph{CoRR}, vol. abs/1703.09452, 2017.
  [Online]. Available: \url{http://arxiv.org/abs/1703.09452}
\BIBentrySTDinterwordspacing

\bibitem{defossez2020real}
A.~Defossez, G.~Synnaeve, and Y.~Adi, ``Real time speech enhancement in the
  waveform domain,'' \emph{arXiv preprint arXiv:2006.12847}, 2020.

\bibitem{time_frequency_domain}
Y.~Xu, J.~Du, L.-R. Dai, and C.-H. Lee, ``An experimental study on speech
  enhancement based on deep neural networks,'' \emph{IEEE Signal Processing
  Letters}, vol.~21, no.~1, pp. 65--68, 2014.

\bibitem{dnsmos_pesq_flaw}
\BIBentryALTinterwordspacing
C.~K.~A. Reddy, E.~Beyrami, J.~Pool, R.~Cutler, S.~Srinivasan, and J.~Gehrke,
  ``A scalable noisy speech dataset and online subjective test framework,''
  2019. [Online]. Available: \url{https://arxiv.org/abs/1909.08050}
\BIBentrySTDinterwordspacing

\bibitem{turian_henry}
\BIBentryALTinterwordspacing
J.~Turian and M.~Henry, ``I'm sorry for your loss: Spectrally-based audio
  distances are bad at pitch,'' 2020. [Online]. Available:
  \url{https://arxiv.org/abs/2012.04572}
\BIBentrySTDinterwordspacing

\bibitem{phoneme_classification_loss}
\BIBentryALTinterwordspacing
P.~Plantinga, D.~Bagchi, and E.~Fosler{-}Lussier, ``Perceptual loss with
  recognition model for single-channel enhancement and robust {ASR},''
  \emph{CoRR}, vol. abs/2112.06068, 2021. [Online]. Available:
  \url{https://arxiv.org/abs/2112.06068}
\BIBentrySTDinterwordspacing

\bibitem{fu2019metricgan}
S.-W. Fu, C.-F. Liao, Y.~Tsao, and S.-D. Lin, ``Metricgan: Generative
  adversarial networks based black-box metric scores optimization for speech
  enhancement,'' in \emph{International Conference on Machine Learning}.\hskip
  1em plus 0.5em minus 0.4em\relax PMLR, 2019, pp. 2031--2041.

\bibitem{metric_RL_SE}
Y.~Koizumi, K.~Niwa, Y.~Hioka, K.~Kobayashi, and Y.~Haneda, ``Dnn-based source
  enhancement self-optimized by reinforcement learning using sound quality
  measurements,'' in \emph{2017 IEEE International Conference on Acoustics,
  Speech and Signal Processing (ICASSP)}, 2017, pp. 81--85.

\bibitem{white_box_perceptual_loss}
J.~M. Martin-Doñas, A.~M. Gomez, J.~A. Gonzalez, and A.~M. Peinado, ``A deep
  learning loss function based on the perceptual evaluation of the speech
  quality,'' \emph{IEEE Signal Processing Letters}, vol.~25, no.~11, pp.
  1680--1684, 2018.

\bibitem{PESQ}
A.~Rix, J.~Beerends, M.~Hollier, and A.~Hekstra, ``Perceptual evaluation of
  speech quality (pesq)-a new method for speech quality assessment of telephone
  networks and codecs,'' in \emph{2001 IEEE International Conference on
  Acoustics, Speech, and Signal Processing. Proceedings (Cat. No.01CH37221)},
  vol.~2, 2001, pp. 749--752 vol.2.

\bibitem{STOI}
C.~H. Taal, R.~C. Hendriks, R.~Heusdens, and J.~Jensen, ``An algorithm for
  intelligibility prediction of time–frequency weighted noisy speech,''
  \emph{IEEE Transactions on Audio, Speech, and Language Processing}, vol.~19,
  no.~7, pp. 2125--2136, 2011.

\bibitem{reddy2021dnsmos}
C.~K. Reddy, V.~Gopal, and R.~Cutler, ``Dnsmos: A non-intrusive perceptual
  objective speech quality metric to evaluate noise suppressors,'' in
  \emph{ICASSP 2021-2021 IEEE International Conference on Acoustics, Speech and
  Signal Processing (ICASSP)}.\hskip 1em plus 0.5em minus 0.4em\relax IEEE,
  2021, pp. 6493--6497.

\bibitem{limitations_of_pesq}
T.~Manjunath, ``Limitations of perceptual evaluation of speech quality on voip
  systems,'' in \emph{2009 IEEE International Symposium on Broadband Multimedia
  Systems and Broadcasting}, 2009, pp. 1--6.

\bibitem{black_box_peq_stoi}
H.~Zhang, X.~Zhang, and G.~Gao, ``Training supervised speech separation system
  to improve stoi and pesq directly,'' in \emph{2018 IEEE International
  Conference on Acoustics, Speech and Signal Processing (ICASSP)}, 2018, pp.
  5374--5378.

\bibitem{other_metric_approach}
Y.~Koizumi, K.~Niwa, Y.~Hioka, K.~Kobayashi, and Y.~Haneda, ``Dnn-based source
  enhancement to increase objective sound quality assessment score,''
  \emph{IEEE/ACM Transactions on Audio, Speech, and Language Processing},
  vol.~26, no.~10, pp. 1780--1792, 2018.

\bibitem{saddler2020speech}
M.~R. Saddler, A.~Francl, J.~Feather, K.~Qian, Y.~Zhang, and J.~H. McDermott,
  ``Speech denoising with auditory models,'' \emph{arXiv preprint
  arXiv:2011.10706}, 2020.

\bibitem{pfpl_paper}
T.-A. Hsieh, C.~Yu, S.-W. Fu, X.~Lu, and Y.~Tsao, ``{Improving Perceptual
  Quality by Phone-Fortified Perceptual Loss Using Wasserstein Distance for
  Speech Enhancement},'' in \emph{Proc. Interspeech 2021}, 2021, pp. 196--200.

\bibitem{seven_ensemble_model}
S.~Kataria, J.~Villalba, and N.~Dehak, ``Perceptual loss based speech denoising
  with an ensemble of audio pattern recognition and self-supervised models,''
  in \emph{ICASSP 2021 - 2021 IEEE International Conference on Acoustics,
  Speech and Signal Processing (ICASSP)}, 2021, pp. 7118--7122.

\bibitem{correlates_breathy}
J.~Hillenbrand, R.~Cleveland, and R.~Erickson, ``Acoustic correlates of breathy
  vocal quality,'' \emph{Journal of speech and hearing research}, vol.~37, pp.
  769--78, 09 1994.

\bibitem{correlates_breathy_rough}
G.~d. Krom, ``Some spectral correlates of pathological breathy and rough voice
  quality for different types of vowel fragments,'' \emph{Journal of Speech,
  Language, and Hearing Research}, vol.~38, no.~4, pp. 794--811, 1995.

\bibitem{KASUYA1986171}
\BIBentryALTinterwordspacing
H.~Kasuya, S.~Ogawa, Y.~Kikuchi, and S.~Ebihara, ``An acoustic analysis of
  pathological voice and its application to the evaluation of laryngeal
  pathology,'' \emph{Speech Communication}. [Online]. Available:
  \url{https://www.sciencedirect.com/science/article/pii/0167639386900063}
\BIBentrySTDinterwordspacing

\bibitem{Gallardo2017PerceivedIS}
L.~F. Gallardo and B.~Weiss, ``Perceived interpersonal speaker attributes and
  their acoustic features,'' 2017.

\bibitem{egemaps}
F.~Eyben, K.~R. Scherer, B.~W. Schuller, J.~Sundberg, E.~André, C.~Busso,
  L.~Y. Devillers, J.~Epps, P.~Laukka, S.~S. Narayanan, and K.~P. Truong, ``The
  geneva minimalistic acoustic parameter set (gemaps) for voice research and
  affective computing,'' \emph{IEEE Transactions on Affective Computing},
  vol.~7, no.~2, pp. 190--202, 2016.

\bibitem{hao2021fullsubnet}
X.~Hao, X.~Su, R.~Horaud, and X.~Li, ``Fullsubnet: a full-band and sub-band
  fusion model for real-time single-channel speech enhancement,'' in
  \emph{ICASSP 2021-2021 IEEE International Conference on Acoustics, Speech and
  Signal Processing (ICASSP)}.\hskip 1em plus 0.5em minus 0.4em\relax IEEE,
  2021, pp. 6633--6637.

\bibitem{reddy2020interspeech}
C.~K. Reddy, V.~Gopal, R.~Cutler, E.~Beyrami, R.~Cheng, H.~Dubey,
  S.~Matusevych, R.~Aichner, A.~Aazami, S.~Braun \emph{et~al.}, ``The
  interspeech 2020 deep noise suppression challenge: Datasets, subjective
  testing framework, and challenge results,'' \emph{arXiv preprint
  arXiv:2005.13981}, 2020.

\bibitem{fu2021metricgan}
S.-W. Fu, C.~Yu, T.-A. Hsieh, P.~Plantinga, M.~Ravanelli, X.~Lu, and Y.~Tsao,
  ``Metricgan+: An improved version of metricgan for speech enhancement,''
  2021.

\bibitem{opensmile}
F.~Eyben, M.~W{\"o}llmer, and B.~Schuller, ``Opensmile: the munich versatile
  and fast open-source audio feature extractor,'' in \emph{Proceedings of the
  18th ACM international conference on Multimedia}, 2010, pp. 1459--1462.

\bibitem{complex_ratio_mask}
D.~S. Williamson, Y.~Wang, and D.~Wang, ``Complex ratio masking for monaural
  speech separation,'' \emph{IEEE/ACM Transactions on Audio, Speech, and
  Language Processing}, vol.~24, no.~3, pp. 483--492, 2016.

\bibitem{graz_university}
G.~Pirker, M.~Wohlmayr, S.~Petrik, and F.~Pernkopf, ``A pitch tracking corpus
  with evaluation on multipitch tracking scenario.'' 01 2011, pp. 1509--1512.

\bibitem{springenberg19_interspeech}
S.~Springenberg, E.~Lakomkin, C.~Weber, and S.~Wermter, ``{Predictive Auxiliary
  Variational Autoencoder for Representation Learning of Global Speech
  Characteristics},'' in \emph{Proc. Interspeech 2019}, 2019, pp. 934--938.

\bibitem{self_attention_pooling}
\BIBentryALTinterwordspacing
P.~Safari, M.~India, and J.~Hernando, ``Self-attention encoding and pooling for
  speaker recognition,'' 2020. [Online]. Available:
  \url{https://arxiv.org/abs/2008.01077}
\BIBentrySTDinterwordspacing

\bibitem{ioffe2015batch}
S.~Ioffe and C.~Szegedy, ``Batch normalization: Accelerating deep network
  training by reducing internal covariate shift,'' in \emph{International
  conference on machine learning}.\hskip 1em plus 0.5em minus 0.4em\relax PMLR,
  2015, pp. 448--456.

\end{thebibliography}


\end{document}